\documentclass[12pt, a4paper]{article}

\usepackage{amssymb, amsmath}  
\usepackage{fullpage}
\usepackage{amsfonts}
\usepackage{graphicx}
\usepackage{mathrsfs}
\usepackage{latexsym}	
\usepackage{oldgerm}
\usepackage{color}
\usepackage{cite}
\usepackage{float}
\usepackage{epstopdf}
\usepackage{cite} 

\newtheorem{definition}{Definition}[section]

\newtheorem{theorem}{Theorem}[section]
\newtheorem{example}{Example}[section]

\newcommand{\bdf}{\begin{definition}}
\newcommand{\edf}{\end{definition}}
\newcommand{\bthm}{\begin{theorem}}
\newcommand{\ethm}{\end{theorem}}
\newcommand{\bex}{\begin{example}}
\newcommand{\eex}{\end{example}}

\newtheorem{prop}{PROPOSITION}[section]
\newcommand{\bprop}{\begin{prop}}
\newcommand{\eprop}{\end{prop}}
\newcommand{\bdis}{\begin{displaymath}}
\newcommand{\edis}{\end{displaymath}}
\newcommand{\beqn}{\begin{equation}}
\newcommand{\eeqn}{\end{equation}}

\newcommand{\bgb}{\begin{gbox}}
\newcommand{\egb}{\end{gbox}}
\newcommand{\bdb}{\begin{darkbox}}
\newcommand{\edb}{\end{darkbox}}

\newcommand{\bsplit}{\begin{split}}
\newcommand{\esplit}{\end{split}}

\usepackage{xcolor}
\usepackage{amsthm}
\usepackage{framed}
\usepackage{authblk}

\colorlet{shadecolor}{gray!15}
\newenvironment{gbox}
  {\begin{shaded}}
  {\end{shaded}}

\title{The Higgs vacuum in the presence of strong gravitational fields: curvature induced electroweak symmetry restoration.} 

\date{\today}

\author{ Paul G. Abel \\
{\small  School of Physics \& Astronomy,} \\
{\small  University of Leicester, Leicester UK.} \\
{\small  LE1 7RH.} \\
{\small {Email:} pga3@le.ac.uk}}

\begin{document}

\maketitle

\begin{abstract}
I investigate the behaviour of the Higgs field in Schwarzschild spacetime by introducing a Planck-suppressed coupling between the Higgs field and the Kretschmann scalar, the leading curvature invariant in Ricci-flat geometries. I show that spacetime curvature modifies the Higgs effective potential, causing the vacuum expectation value to decrease continuously with increasing curvature. A critical curvature is derived at which the Higgs vacuum expectation value vanishes, signalling a curvature-induced restoration of electroweak symmetry. For Schwarzschild black holes, we obtain analytical expressions for the critical curvature, the corresponding phase-transition radius, and the Planck-curvature radius that marks the onset of quantum-gravity effects. These results identify a self-consistent region inside the event horizon in which electroweak symmetry is restored while the effective description remains applicable. The model provides a simple analytical framework linking strong gravitational fields to the Higgs vacuum through a single dimensionless curvature-Higgs coupling.

\end{abstract}

\section{Introduction}
The discovery of the Higgs boson completed the Standard Model of particle physics and confirmed the mechanism responsible for electroweak symmetry breaking. Through spontaneous symmetry breaking, the Higgs field acquires a non-zero vacuum expectation value, $v \simeq 246,\mathrm{GeV}$ \cite{higgsvacexpvalue}, generating the masses of the weak gauge bosons and fermions while leaving the photon massless. Although the Higgs mechanism has been extensively studied in flat spacetime and at finite temperature, considerably less is known about its behaviour in the presence of strong gravitational fields, where spacetime curvature may itself modify the structure of the Higgs vacuum. A number of interesting results have shown that gravitational interactions can have an effect on the lifetime of the Higgs field when quantum gravity corrections are made \cite{PhysRevLett.115.071303}, and that vacuum fluctuations of the Higgs induce a gravitational collapse of the vacuum \cite{PhysRevD.98.123509}
\\

Quantum field theory in curved spacetime provides the natural framework for studying quantum fields propagating on a classical curved spacetime background. In this approach, gravity modifies the dynamics of matter fields through the local spacetime geometry while the gravitational field itself remains classical.  Early work established that spacetime curvature can play an important role on quantum mechanical effects near the event horizon of a black hole \cite{Hawking, Birrell_Davies_1982, qftcurved} and continues to be a very active area of research today. 
\\

Curvature corrections to scalar fields are therefore expected to arise naturally, either through non-minimal couplings or through higher-dimensional operators generated in an effective field theory describing physics below some ultraviolet completion. Understanding how these corrections influence the Higgs vacuum is important not only for black-hole physics but also for the behaviour of matter in the strong-field regime of gravity. 
\\

Many of the previous studies have considered the familiar non-minimal interaction $\xi RH^2$, where $R$ is the Ricci scalar, for example \cite{GIALAMAS2023137885}. Such couplings play a central role in Higgs inflation and in scalar field dynamics in cosmological spacetimes. However, the Schwarzschild solution is Ricci-flat, and so $R=0$, and $R_{\mu\nu}=0$.  Consequently the leading Ricci-scalar and Ricci-tensor couplings vanish identically. They therefore cannot modify the Higgs vacuum in the exterior or interior of an uncharged Schwarzschild black hole. 
\\

Any gravitational modification of the Higgs sector in a Ricci-flat geometry must instead arise from higher-order curvature invariants.  A simple non-vanishing invariant in Schwarzschild spacetime is the Kretschmann scalar,
\beqn
K=R_{\mu\nu\rho\sigma}R^{\mu\nu\rho\sigma},
\eeqn
which measures the magnitude of the spacetime curvature independently of the choice of coordinates and increases monotonically towards the central singularity. Since $K$ is the leading scalar curvature invariant that remains non-zero in Ricci-flat spacetimes, it provides the natural quantity through which gravity can couple to the Higgs field. Motivated by effective field theory, we therefore introduce the interaction
\beqn
\mathcal L_{\rm Grav}=-\xi_K \ell_{\rm Pl}^2K H^2,
\eeqn
where $\xi_K$ is a dimensionless Kretschmann curvature coupling and $\ell_{\rm Pl}$ is the Planck length. The factor $\ell_{\rm Pl}^2$ provides the required dimensional suppression expected for the leading higher-curvature correction generated by quantum-gravitational physics, while the dimensionless coefficient $\xi_K$ parametrises the strength of the interaction. This operator represents the lowest-order curvature-Higgs coupling capable of influencing the Higgs vacuum in a Ricci-flat spacetime.  Other researchers have also used the Kretschmann scalar, for example in the investigation of Higgs induced spectroscopic shifts in a strong gravitational field \cite{PhysRevD.82.065008}
\\

The physical interpretation of this interaction is straightforward. The Kretschmann scalar acts as a local measure of gravitational field strength and therefore contributes an effective curvature-dependent correction to the Higgs mass parameter. As the curvature increases towards the black-hole interior, the Higgs effective potential is progressively modified, allowing gravity itself to influence the stability of the electroweak vacuum. Unlike thermal symmetry restoration in the early Universe, where temperature is the control parameter, the present mechanism is driven entirely by spacetime curvature. The Kretschmann scalar therefore plays the role of a geometric order parameter governing the evolution of the Higgs vacuum.
\\

In this paper I investigate the consequences of this curvature-Higgs interaction in Schwarzschild spacetime. I derive the curvature-modified Higgs effective potential and obtain an analytic expression for the curvature-dependent vacuum expectation value. This leads to a critical curvature at which the Higgs vacuum expectation value vanishes, identifying a curvature-induced restoration of electroweak symmetry. Applying the formalism to the Schwarzschild geometry, we derive analytical expressions for the corresponding phase-transition radius and for the radius at which Planck-scale curvature is reached. These characteristic radii divide the black hole interior into distinct physical regions in which electroweak symmetry is broken, curvature-restored, or beyond the regime where the classical effective description is expected to remain valid.
\\

The results suggest that the leading Planck-suppressed coupling between the Higgs field and the Kretschmann scalar provides a simple analytical mechanism by which strong gravitational fields can modify the electroweak vacuum in Ricci-flat spacetimes. The resulting curvature-induced phase transition establishes a direct connection between black hole geometry and the Higgs sector and provides a framework for investigating the interplay between particle physics and strong gravitational fields.

\section{The Higgs vacuum in the presence of strong gravity}

We start by taking our Lagrangian for the system to be
\beqn
\mathcal L =\sqrt{-g}\left[ \frac 1 2 \nabla_\mu H\nabla^\mu H-\lambda(H^2-v^2)^2-\xi_K \ell^2_{Pl}KH^2 \right]
\eeqn
where $H$ is the Higgs scalar field, so the covariant derivative in the kinetic term reduces to $\nabla_\mu H\nabla^\mu H=\partial_\mu H\partial^\mu H$. The constant $\lambda$ represents the strength of the Higgs-self interaction, while $v$ is the Higgs expectation value.  
\\

The last term in the Lagrangian represents the presence of the strong gravitational interaction as discussed previously. The Kretschmann scalar $K$, for the Schwarzschild spacetime has the value
\beqn
K=\frac{48G^2M^2}{r^6}.
\label{KS}
\eeqn

Within an effective field theory, general covariance permits higher-dimensional operators coupling the Higgs field to spacetime curvature.  In Ricci-flat spacetimes, the leading non-vanishing scalar interaction is the dimension 6 operator $KH^2/\Lambda^2$, which therefore provides the natural lowest-order correction to the Standard Model Higgs Sector. 
\\

The new effective potential is 
\beqn
V(H)=\lambda(H^2-v^2)^2+\xi_K \ell^2_{Pl}KH^2
\label{Vf}
\eeqn
We see from (\ref{Vf}) that only the $H^2$ term is affected by gravity, thus indicating already that we might expect gravity to alter the Higgs mass.  
\\

To obtain the curvature-dependent Higgs vacuum from the effective potential, I work in the local effective-potential approximation.  This assumes that the Higgs field varies slowly on the length scale over which the Schwarzschild curvature changes, so that the gradient contributions from $\nabla_\mu H\nabla^\mu H$ are subleading compared with the potential terms.  The relevant curvature length scale is $\L_K\sim |K/\partial_rK|=r/6$, while the local Higgs correlation length is $m_{\rm eff}^{-1}$, where $m_{\rm eff}^2=V''(H_0)$ in the broken phase.  
\\

The neglect of the kinetic terms is therefore justified whenever $m_{\rm eff}L_K\gg 1$.  This condition is expected to hold throughout the adiabatic bulk of the effective-field theory regime but of course it can fail in a narrow region near the critical curvature where $m_{\rm eff}\rightarrow 0$, and also near the Planckian curvature, where the effective description itself breaks down.  The algebraic minimisation of the potential $V(H,K$) should therefore be interpreted as determine the local equilibrium value of the Higgs field away from these transition and cut-off regions.
\\

We now derive the new Higgs vacuum in the presence of gravity:
\bdis
\frac{dV}{dH}= 2H\left[2\lambda(H^2-v^2) +\xi_K \ell_{\rm Pl}^2K\right]
\edis
and so solving $V'(H)=0$ gives
\beqn
H_0^2=v^2-\frac{\xi_K \ell_{\rm Pl}^2 K}{2\lambda}
\label{H0}
\eeqn
which we will write as
\beqn
v_{\rm eff}=\sqrt{v^2-\frac{\xi_K \ell_{\rm Pl}^2 K}{2\lambda}}
\label{Higgs vac}
\eeqn
which clearly reduces to the standard Higgs vacuum if $K=0$.  For $v_{\rm eff}=0$, we see that
\beqn
v^2=\frac{\xi_K\ell_{\rm Pl}^2K_{\rm crit}}{2\lambda}
\eeqn
so this gives a critical value for the curvature term:
\beqn
K_{\rm crit}=\frac{2\lambda v^2}{\xi_K\ell_{\rm Pl}^2}
\label{KC}
\eeqn
In the case of Schwarzschild spacetime, we have by (\ref{KS}) that if $K=K_{\rm crit}$
\beqn
K_{\rm crit}=\frac{48G^2M^2}{r^6}
\label{KC}
\eeqn
which gives 
\beqn
r_{\rm crit}=\left[ \frac{24\xi_K \ell_{\rm Pl}^2G^2M^2}{\lambda v^2} \right]^{\frac{1}{6}} 
\label{rc}
\eeqn

\section{Stability of the new curvature induced Higgs vacuum}

We now examine the stability of the new vacuum in the presence of gravity.  This can be determined by looking at the second derivative evaluated at the minimum $H_0$. We have that
\beqn
V''(H_0)=12\lambda H_0^2-4\lambda v^2+2\xi_K\ell_{\rm Pl}^2K
\eeqn
and so by (\ref{H0}) we have that
\beqn
V''(H_0)=4(2\lambda v^2-\xi_K\ell_{\rm Pl}^2K).
\eeqn
Using our expression for $K_{\rm crit}$ in (\ref{KC}) we have that
\beqn
V''(H_0)=4\xi_K\ell_{\rm Pl}^2(K_{\rm crit}-K).
\eeqn

This expression immediately determines the stability of the broken electroweak vacuum.
For $K<K_{\rm crit}$, the second derivative is positive, showing that the two degenerate minima at $H=\pm H_0$ are stable and that electroweak symmetry remains spontaneously broken.
\\

At the critical curvature, $K=K_{\rm crit}$, the second derivative vanishes, $V''(H_0)=0$,
indicating that the curvature of the potential disappears and the two broken-symmetry minima merge continuously with the symmetric configuration at $H=0$. This identifies the curvature-induced electroweak phase transition.
\\

It should be noted that since $V''(H_0) \rightarrow 0$ at the critical curvature, the Higgs correlation length diverges and the local-potential approximation is not expected to resolve the detailed structure of the transition layer.  Including the kinetic terms would smooth the radial interpolation of the Higgs field across $r_{\rm crit}$, but it does not alter the mean-field criterion that the curvature-dependent quadratic term changes sign at $K=K_{\rm crit}$.
\\

It is important to emphasise that the expression for $V''(H_0)$ is physically meaningful only while the broken-symmetry solution exists. Once $K>K_{\rm crit}$
the quantity
\bdis
H_0^2=v^2-\frac{\xi_K\ell_{\rm Pl}^2K}{2\lambda}
\edis
becomes negative, so the stationary point $H_0$ is no longer a real solution of the field equations. Consequently, the continuation of $V''(H_0)$ into the region $K>K_{\rm crit}$ has no physical interpretation and cannot be used to determine vacuum stability.
For curvatures exceeding the critical value, the only physical stationary point is $H=0.$
\\

The stability of the symmetric vacuum is therefore determined by
\bdis
2\xi_K\ell_{\rm Pl}^2
\left(K-K_{\rm crit}\right).
\edis
Hence, $V''(0)>0$ $(K>K_{\rm crit})$, demonstrating that the symmetric vacuum becomes the unique stable minimum once the critical curvature is exceeded.
The stability analysis therefore provides a complete description of the curvature-induced phase transition. Below the critical curvature the stable vacuum is characterised by a non-zero Higgs vacuum expectation value, while above the critical curvature the broken vacuum ceases to exist and the stable solution becomes $H=0$. The transfer of stability from the broken vacuum to the symmetric vacuum occurs continuously at
\bdis
K=K_{\rm crit},
\edis
confirming that the Kretschmann scalar acts as the geometric control parameter governing curvature-induced electroweak symmetry restoration.

\section{Conclusions}

In this work, we have investigated the influence of strong spacetime curvature on the Higgs vacuum by introducing a Planck-suppressed coupling between the Higgs field and the Kretschmann scalar. Since the Schwarzschild spacetime is Ricci-flat, the commonly studied non-minimal coupling $\xi RH^2$ cannot modify the Higgs sector. The Kretschmann scalar therefore provides the leading non-vanishing curvature invariant through which gravity can influence electroweak symmetry breaking.
\\

Using the curvature-modified Higgs potential, we derived an analytic expression for the curvature-dependent vacuum expectation value,
\bdis
v_{\rm eff}^2=v^2-\frac{\xi_K\ell_{\rm Pl}^2K}{2\lambda},
\edis
demonstrating that increasing spacetime curvature continuously suppresses the Higgs vacuum expectation value.
The condition
\bdis
K_{\rm crit}=\frac{2\lambda v^2}
{\xi_K\ell_{\rm Pl}^2}
\edis
defines a critical curvature at which the Higgs vacuum expectation value vanishes. Analysis of the second derivative of the effective potential shows that the broken electroweak vacuum is stable for $K<K_{\rm crit}$, while the symmetric vacuum becomes stable for $K>K_{\rm crit}.$
At $K=K_{\rm crit}$,
the two phases meet continuously, identifying a curvature-induced second-order electroweak phase transition.
\\

Applying these results to the Schwarzschild geometry, we obtained closed-form expressions for the phase-transition radius,
\bdis
r_{\rm crit}=\left(
\frac{24\xi_K\ell_{\rm Pl}^2G^2M^2}
{\lambda v^2}
\right)^{1/6},
\edis
and for the Planck-curvature radius,
\bdis
r_{\rm EFT}=\left(
48G^2M^2\ell_{\rm Pl}^4
\right)^{1/6},
\edis
which marks the onset of the regime in which quantum-gravitational corrections are expected to become significant.
These results divide the Schwarzschild interior into three distinct physical regions. This division assumes the ordering
\beqn
r_{\rm EFT}<r_{\rm crit}<r_h,
\eeqn
so that there exists a finite region in which electroweak symmetry is restored while the effective-field-theory description remains valid.  This ordering is not automatic for arbitrary values of the dimensionless coupling $\xi_K$, but instead imposes a corresponding range on $\xi_K$.  In what follows we assume that $\xi_K$ lies in this range.
\\

For
\beqn
r_{\rm crit}<r<r_h,
\eeqn
electroweak symmetry remains spontaneously broken despite the strong gravitational field. For
\beqn
r_{\rm EFT}<r<r_{\rm crit},
\eeqn
spacetime curvature restores electroweak symmetry while the effective description remains valid. Finally, for $r<r_{\rm EFT}$,
the spacetime curvature approaches the Planck scale, and the classical effective theory is no longer expected to provide a complete description.
\\

The principal result of this work is that a single Planck-suppressed Kretschmann-Higgs interaction is sufficient to generate an analytic curvature-induced electroweak phase transition in a Ricci-flat black hole spacetime. The Kretschmann scalar therefore acts as a geometric control parameter for electroweak symmetry breaking, providing a direct link between strong gravitational fields and the structure of the Higgs vacuum.
It should be noted that this model is relatively simple; only the tree-level Higgs potential is used, no Coleman or Weinberg corrections are included and finally the back-reaction on the metric has been ignored.
\\

Nonetheless, this framework offers a simple and self-consistent extension of the Standard Model in curved spacetime and suggests that black hole interiors are good laboratories to study the dynamics of the Higgs field in more general strong-gravity backgrounds.

\paragraph{Acknowledgement:}  The author would like to thank Derek Raine, Professor emeritus of the University of Leicester for a number of useful discussions during the time this work was undertaken.

\newpage
\bibliographystyle{elsarticle-num}  
\bibliography{references}  

@article{higgsvacexpvalue,
	author={C. Amsler and M. Doser and M. Antonelli and D.M. Asner and K.S. Babu and H. Baer and others},
	title={Review of Particle Physics},	
	journal={Physics Letters B},
	volume={667},
	number={1},
	pages={1-6},
	year={2008}
}

@article{PhysRevLett.115.071303,
	author={Philipp Burda and Ruth Gregory and Ian G. Moss},
	title={Gravity and the Stability of the Higgs Vacuum},
	journal={Phys. Rev. Lett.},
	volume={115},
	number={7},
	pages={071303},
	year={2015}
}

@article{PhysRevD.98.123509,
	author={Kazunori Kohri and Hiroki Matsui},
	title={Electroweak vacuum collapse induced by vacuum fluctuations of the Higgs field around evaporating black holes},
	journal={Phys. Rev. D},
	volume={98},
	issue={12},
	pages={123509},
	numpages={11},
	year={2018}
}

@article{Hawking,
	author={Stephen W. Hawking},
	title={Particle creation by black holes},
	journal={Comm. Math. Phys.},
	volume={43},
	number={3},
	pages={199-220},
	year={1975}
}

@book{Birrell_Davies_1982, place={Cambridge}, series={Cambridge Monographs on Mathematical Physics}, title={Quantum Fields in Curved Space}, publisher={Cambridge University Press}, author={Birrell, N. D. and Davies, P. C. W.}, year={1982}, collection={Cambridge Monographs on Mathematical Physics}}

@article{qftcurved,
	author={Bryce DeWitt},
	title={Quantum field Theory in Curved Spacetime},
	journal={{Physics Reports}},
	volume={19},
	number={6},
	pages={295-357},
	year={1975},
}

@article{GIALAMAS2023137885,
	author={Ioannis D. Gialamas and Alexandros Karam and Thomas D. Pappas},
	title={Gravitational corrections to electroweak vacuum decay: metric vs. Palatini},
	journal={Physics Letters B},
	volume={840},	
	pages={137885},
	year={2023},
}

@article{PhysRevD.82.065008,
	title={Higgs-induced spectroscopic shifts near strong gravity sources},
	author={Onofrio, Roberto},
	journal={Phys. Rev. D},
	volume={82},
	number={6},
	pages={065008},
	year={2010},
	publisher={American Physical Society},
}

\end{document}